# Towards a mechanism of rattler coupling in the β-pyrochlores $AOs_2O_6$ (A = K, Rb, Cs)

*Elvis Shoko[a)], Vanessa K Peterson, and Gordon J Kearley*

Australian Nuclear Science and Technology Organisation, Locked Bag 2001, Kirrawee DC, NSW 2232, Australia

## Abstract

We have applied *ab initio* molecular dynamics (MD) simulations to study metal-metal coupling on the alkali-metal sublattice in the β-pyrochlore osmates, $AOs_2O_6$ (A = K, Rb, Cs) at 300 K. We find that the alkali-metal atoms (rattlers) couple to each other more strongly than they couple to the cage atoms, and that, at 300 K, this coupling is strongest for Cs. We show that this coupling controls the dominant dynamics in the rattling of these atoms. We provide preliminary evidence that the rattlers couple to each other primarily through the $T_{2g}$ mode whereas their coupling to the cage modes occurs through the $T_{1u}$ mode. Rattler coupling through the $T_{2g}$ mode provides insight into the trend in spectral broadening from Cs to K. The spectral broadening is inversely proportional to the strength of the dynamical correlations on the alkali-metal sublattice which in turn depend on the atomic size of the rattler, decreasing from Cs to K. Thus, the broadest spectrum exhibited by the K is partly a consequence of the small size of this rattler which permits a greater range of motions involving combinations of both correlated and anti-correlated dynamics. We emphasize that the identification of the somewhat distinct roles of the $T_{1u}$ and $T_{2g}$ modes in rattler coupling reported in this work is a significant step towards a complete fundamental mechanism of rattler dynamical coupling in these osmates. We believe that such a mechanism will have profound implications for a broad class of cage compounds, including clathrates and skutterudites.

---

[a)] Author to whom correspondence should be addressed. Electronic mail: elvis.shoko@gmail.com. Telephone: +61-41-506-4823

# I. INTRODUCTION

Rattling dynamics represent a phenomenon of great interest in understanding superconductivity in cage compounds, e.g., the beta-pyrochlore osmates,[1] and some vanadium intermetallics[2] while in clathrates[3] and skutterudites,[4] the interest derives from their potential role in thermoelectric properties. The mechanism of rattling dynamics, and in particular, the nature of the interaction between the rattler and its local environment, has been the subject of intense investigation. For instance, in the clathrates[3] and skutterudites,[4] the issue of whether the rattler reduces the lattice thermal conductivity by resonant scattering or localized-mode coupling to the lattice modes is still under investigation.[5, 6] In this study, we focus on the β-pyrochlore osmates, $AOs_2O_6$ (A = K, Rb, Cs) and apply *ab initio* molecular dynamics (MD) to examine the details of rattler coupling in these materials at 300 K. The pyrochlore osmates crystallize in space-group $Fd\bar{3}m$,[7] where $A$, Os, and O, occupy the 8$b$, 16$c$, and 48$f$ sites, respectively. The transition metal cation, Os, is octahedrally coordinated by O and the $OsO_6$ octahedra share corners. The key structural feature of these compounds relevant to alkali-metal rattling is the existence of alkali-metal atoms encapsulated in an $Os_{12}O_{18}$ cage presenting a large volume in which the alkali metal can undergo unusual dynamics. These structural elements are illustrated in FIG. 1 where the fourfold degenerate 32$e$ site which some reports[8-10] suggest to be the site occupied by the K atoms at temperatures ≤ 100 K is also included.

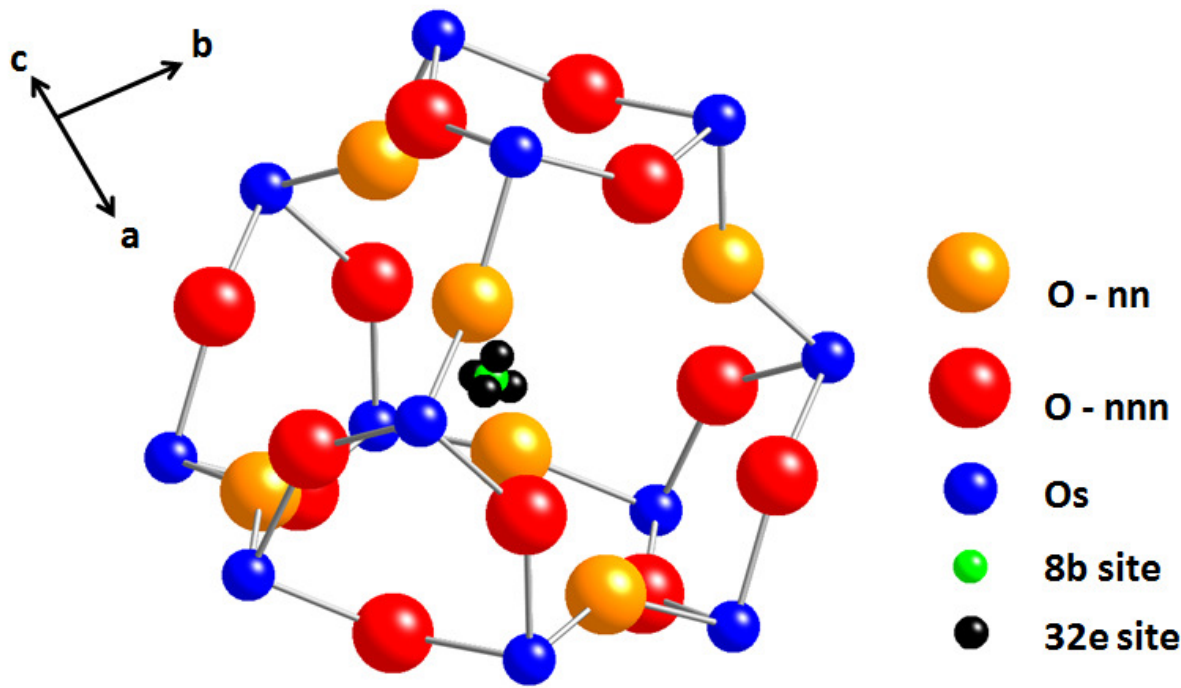

FIG. 1. Cage structure in the defect-pyrochlore osmates. The cage consists of six nearest-neighbor O atoms (O - nn) octahedrally arranged around the cage center followed by twelve next-nearest-neighbor O atoms (O - nnn). There are twelve Os atoms per cage as shown. The two types of alkali-metal atom sites are also shown: the 8*b* site at the center of the cage and the 32*e* site which is off-center and of fourfold degeneracy. An alkali metal at these sites is weakly bonded to its cage allowing it to undergo unusual dynamics in the large volume. The coordinate axes refer to the lattice vectors of the conventional cubic unit cell.

Different experimental approaches have been applied to investigate rattling in the β-pyrochlore osmates including detailed structural characterizations, specific heat measurements, NMR, Raman scattering, and inelastic neutron scattering (INS), and more recently, *ab initio* MD simulations.[11, 12] Detailed structural analyses revealed relatively large atomic displacement parameters (ADPs) for all the alkali metals,[13, 14] with K exhibiting the largest ADPs indicating weak bonding to the cage and thus providing indirect evidence for rattling dynamics of the alkali metals. Einstein terms were required to fit specific heat data to the phenomenological Debye model leading to the identification of the alkali metals as Einstein oscillators (rattlers) exhibiting localized rattling modes.[15-18] NMR relaxation rates revealed an anomalous relaxation of the K but not Rb atoms[19] which was governed by the electric-field gradient rather than the magnetic fluctuations. This type of relaxation, described by collisional damping,[20, 21] was attributed to a highly anharmonic rattling of K ions consistent with known rattling vibrations in other cage compounds.[22] Further evidence of the anharmonicity of alkali-metal rattling was the linear temperature dependence of the energy of the $T_{2g}$ mode from 4 to 300 K obtained from Raman scattering.[23, 24] $KOs_2O_6$ showed the strongest temperature dependence with the energy of the $T_{2g}$ mode changing by 1.1 meV in that temperature range. These anharmonicities resembled those found for rattling guest atoms in clathrates[25] and skutterudites.[26] Furthermore, the line width of the $T_{2g}$ mode increased from Cs to K reaching about 1 meV at 4 K attributed to stronger rattling characteristics for K.[24] In addition to the already known Raman-active $T_{2g}$ mode, INS provided the first direct observation of the $T_{1u}$ mode (not Raman-active) of the alkali metals which is lower than the $T_{2g}$ mode.[27, 28] Particularly important was the observation of a more complex anharmonic behavior of the split band for $KOs_2O_6$ at 300 K ($T_{1u}$ at 5.5 meV + $T_{2g}$ at 6.8 meV) with decreasing temperature than was understood from Raman data alone. Significant softening of the $T_{1u}$ mode was observed with decreasing temperature, peaking at 3.4 meV at 1.5 K, the lowest temperature in the experiment. In contrast, the energy of the $T_{2g}$ mode remained almost constant with cooling while the intensity of the mode decreased significantly, virtually vanishing at 1.5 K, presumably due to damping. Overall, the experimental evidence showed that the vibrational dynamics of the K atom in $KOs_2O_6$ differ significantly from those of Rb and Cs in their respective pyrochlores[1] and a complex low-energy signature was recently observed using inelastic neutron scattering (INS).[29]

Significant theoretical work has also been carried out to understand alkali-metal rattling in the β-pyrochlore osmates.[11, 30-32] We restrict this review to work focusing on the issue of metal-metal coupling on the alkali-metal sublattice. Inspired by their finding of the instability of the symmetric ($A_g$) K phonon mode,[33] Kunes and Pickett constructed an effective ionic Hamiltonian consisting only of terms from the alkali-metal sublattice with the O-Os framework only providing a fixed charge background:[32]

$$\widehat{H} = \sum_i \left[ \frac{p_i^2}{2M} + P_e(\xi_i) + P_o(\xi_i) Y_2^3(\xi_i) \right] + \sum_{i>j} W_{ij}(\xi_i, \xi_j) \qquad (1)$$

where the first term is the on-site energy and the second term is the interaction energy which is restricted to nearest-neighbor (nn) pair interactions ($W_{ij}$) on the alkali-metal sublattice. The on-site Hamiltonian is made up of the kinetic energy of the ion ($p_i$ = momentum, M=mass), and the potential energy parameterized in the displacement, $\xi_i$, of the ion. $P_e$ and $P_o$ are polynomial radial functions obtained by fitting *ab initio* total energies as a function of displacements of a single ion along the (111) direction with a polynomial and taking its even and odd parts respectively. Only the lowest even (*l*=0) and odd (*l*=3; *l*=1 vanishes by symmetry) spherical harmonics were used. Solution of the on-site Hamiltonian gave, for K, a quasi-four-fold degenerate ground state[34] corresponding to the 32*e* sites of this ion.[8] When the full Hamiltonian (Eq. 3) was solved in the basis of the product states of the on-site Hamiltonian, non-negligible screened Coulomb interaction between the ions was obtained only for the K atom[9, 32] apparently driven by the hopping of this ion between the 32*e* sites.[34] We note that whether the K atoms occupy the 32*e* or 8*b* sites appears to still be in dispute.[35-37] When the interaction Hamiltonian in Eq. 3 was reformulated in the basis of "bonds" between the interacting K ions, a lattice Hamiltonian equivalent to a four-state Potts model[38] with a constraint was obtained. The constraint, which prohibited population of higher energy states, led to frustration of the K ion dynamics on the diamond sublattice.[9] This work concluded that metal-metal coupling is important for the K, but not for the Rb nor Cs ion dynamics. The model was subsequently extended by Hattori and Tsunetsugu to include five-center (8*b* and 32*e* sites) hoping of the K ion and Coulomb interactions up to the next-nearest neighbors.[10]

In this *ab initio* MD study of the alkali-metal dynamics of the β-osmates at 300 K, we first investigate metal-metal coupling on the alkali-metal sublattice by calculating the Pearson correlation coefficients for the various atoms from their respective trajectories. The evidence of strong alkali-metal coupling on the alkali-metal sublattice that we find is discussed. We then examine the roles of the $T_{1u}$ and $T_{2g}$ mode in rattler coupling and show that the weaker $T_{1u}$ mode is responsible for coupling the rattler to its cage while the stronger $T_{2g}$ mode couples it to other rattlers. Lastly, we discuss the trend in spectral broadening from Cs to K in terms of the strong alkali-metal sublattice coupling.

## II. COMPUTATIONAL DETAILS

We performed low-precision *ab initio* MD simulations using the Vienna Ab initio Simulation Package (VASP)[39, 40] on the conventional cubic unit cell of the β-pyrochlore osmates, $AOs_2O_6$ (A = K, Rb, Cs) consisting of 72 atoms, FIG 2. The Projector-Augmented Wave (PAW) method[41, 42] implemented in the Vienna Ab-initio Simulation Package (VASP)[39, 40] was employed in this study, and for the exchange-correlation potential, the generalized gradient approximation with the Perdew, Burke, and Ernzerhof (GGA-PBE) functional was used.[43, 44] Calculations were performed at the Gamma point with a plane wave cutoff energy of 300 eV. Starting from the experimental atomic positions for the K,[13] Rb,[14] and Cs[18] pyrochlores, the structures were first converged to < 0.001 eV/Å total forces per atom followed by equilibration for 4 ps at 300 K and microcanonical ensemble (NVE) production runs of 36 ps were then performed.

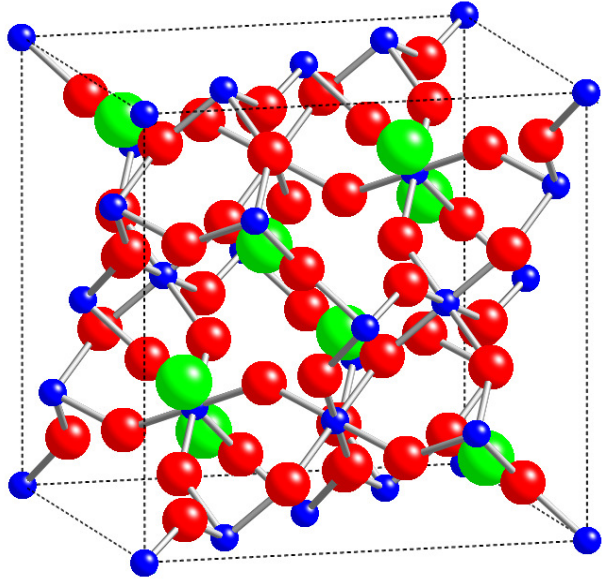

**FIG. 2. The conventional cubic unit cell of the defect pyrochlore osmates used in the MD simulations. The cell consists of 72 atoms shown here as large black (alkali-metal atoms), medium-size red (O atoms), and small blue (Os atoms) spheres.**

For each pyrochlore, two types of calculations were performed: (1) the standard MD simulation where all the atoms in the unit cell were allowed to move in response to the forces acting on them, and (2) the 'static' lattice simulations in which atoms forming the cage framework were kept fixed at their initial positions throughout the simulation. The static simulations are similar to the effective ionic Hamiltonian model of Kunes and Pickett[9] where the cage framework provides a fixed charge background. For the $KOs_2O_6$ case, an additional simulation was performed with the masses of the cage framework atoms increased fourfold – 'heavy' lattice.

## III. RESULTS AND DISCUSSION

It is important to validate MD trajectories against experimental data to ensure that they correctly describe the physical dynamics before extracting any physical quantities from them. Whereas the results from the standard simulations have been validated against experimental INS spectra,[11] 'static' lattice simulations are unphysical and therefore cannot be directly compared to experiment. However, since only minor modifications to the validated standard simulations are involved, they are expected to correctly represent the real dynamics in such a system if it was possible to perform an experiment. In Sec. III.A we analyze the MD trajectories for evidence of metal-metal sublattice coupling and show that significant coupling indeed occurs but differs from that previously reported[9] in that it increases with atomic size. Section III.B examines the nature of rattler dynamical coupling on the basis of the two main vibrational modes as observed experimentally, i.e., the $T_{1u}$ and $T_{2g}$ modes. By comparing between the standard and static lattice simulations, the role of each of these modes in the different pyrochlores is clarified. Lastly, in Sec. III.C, we discuss the broadening of the $T_{2g}$ mode from Cs to K in terms of the metal-metal sublattice coupling.

### A. Evidence for strong coupling on the alkali metal sublattice

Previous studies[9, 32, 34] have suggested that the K atoms strongly couple to each other through a Coulombic interaction on the diamond sublattice which they occupy while remaining weakly coupled to the cages in which they are encapsulated. A four-state Potts model was then proposed and used to describe the correlation on the K sublattice and this accounted for frustrated ion-dynamics on this sublattice.[9] From our MD results, we have probed for evidence of coupling on the K sublattice and extended the investigation to also include the Rb and Cs analogues. For this discussion, we will refer to the coupling between the alkali-metal atoms as A-A coupling (A = K, Rb, Cs). In FIG. 3-5, we plot the Pearson's correlation coefficient between each of the 8 K (or Rb, Cs) atoms in the MD simulation cell and the rest of the other atoms as a function of the distance between them. For each pair of atoms, we calculate the dynamical correlation matrix in the principal axes basis and in FIG. 3-5 we plot the absolute value of the matrix element with the largest magnitude. Double counting of pairs occurs in this procedure but this doesn't matter as each double count maps onto the same point in FIG. 3-5. We also note that correlation coefficients for separation distances > 5 Å may contain artifacts of the periodic boundary conditions (PBCs) applied to the simulation. However, the structural similarity between these pyrochlores suggests that these effects would be similar for all simulations. Thus at these distances, we are only interested in making comparison across the series rather than the absolute values of the individual correlations. Considering only correlations within 5 Å, FIG. 3-5 show that, despite being farther away, the alkali metals couple to each other more strongly than they do to any other atoms for each pyrochlore.

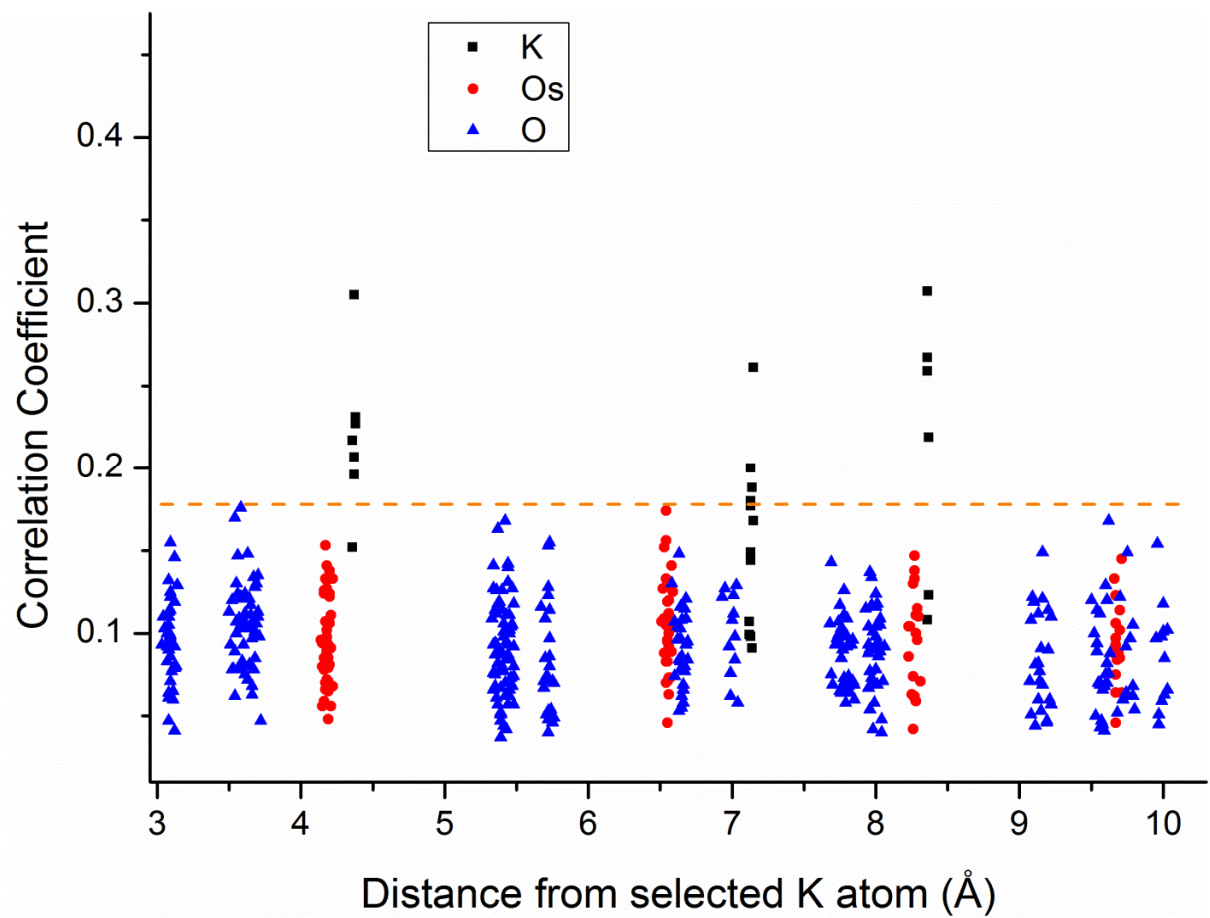

FIG. 3. Absolute correlation coefficients between each K atom and each of the other atoms in the MD simulation cell plotted as a function of the distance between the atoms. The MD data shows that the K atoms couple to each other more than they couple to the other atoms in the rest of the lattice and that the coupling is long-range. The dashed horizontal line marks the maximum for the correlations from the cage atoms. Correlations at distances > 5 Å may be affected by periodic boundary conditions applied in the simulation and are included here only for comparisons across the K-Cs series (see text).

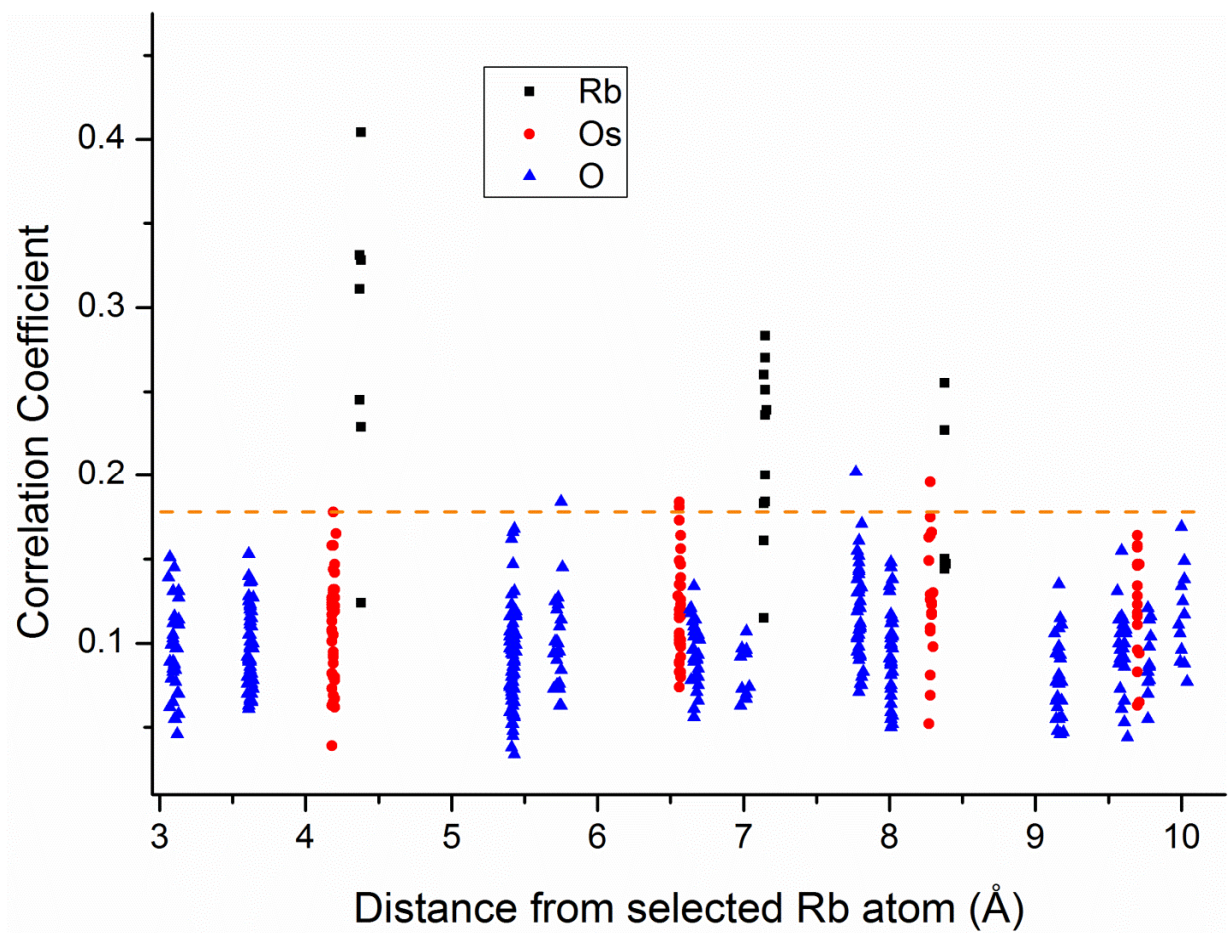

FIG. 4. Absolute correlation coefficients between each Rb atom and each of the other atoms in the MD simulation cell plotted as a function of the distance between the atoms. The MD data shows that the Rb atoms couple to each other more than they couple to the other atoms in the rest of the lattice and that the coupling is long-range. The dashed horizontal line marks the maximum for the correlations from the cage atoms for $kOs_2O_6$. Correlations at distances > 5 Å may be affected by periodic boundary conditions applied in the simulation and are included here only for comparisons across the K-Cs series (see text).

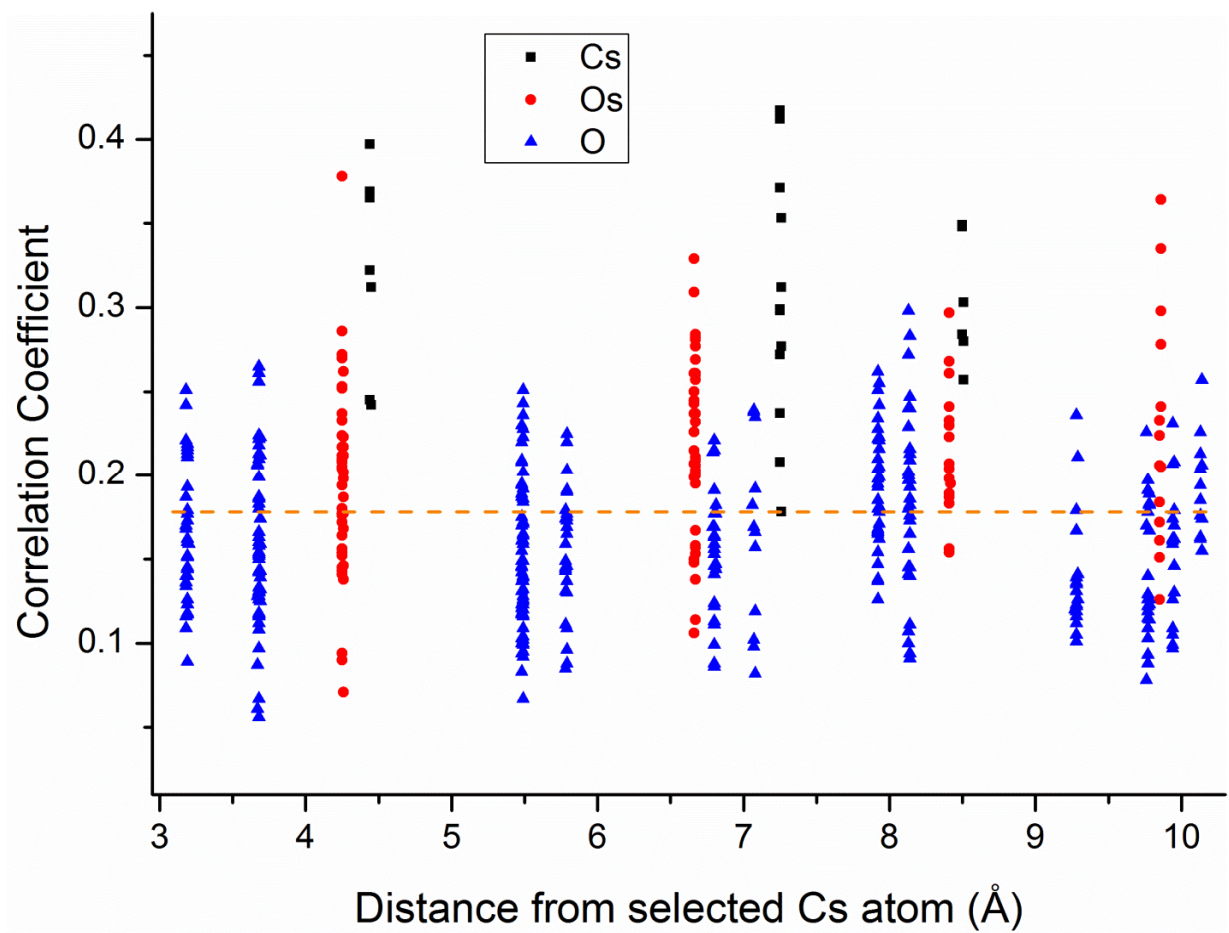

FIG. 5. Absolute correlation coefficients between each Cs atom and each of the other atoms in the MD simulation cell plotted as a function of the distance between the atoms. The MD data shows that the Cs atoms couple to each other more than they couple to the other atoms in the rest of the lattice and that the coupling is long-range. The dashed horizontal line marks the maximum for the correlations from the cage atoms for $kOs_2O_6$ indicating that the corresponding Cs correlations are larger. Correlations at distances > 5 Å may be affected by periodic boundary conditions applied in the simulation and are included here only for comparisons across the K-Cs series (see text).

In FIG. 3-5, we plotted a horizontal dashed line to mark the maximum for the correlations of the $KOs_2O_6$ cage atoms to show that the Cs atoms are more correlated to their cage atoms compared to both K and Rb. However, the comparison between K and Rb is not clear-cut; considering only the first coordination shell, the K exhibits slightly more, and less, correlated O and Os atoms, respectively, than Rb. Thus no particular trend appears to emerge with atomic size of the alkali metal for the A-cage correlations in the K-Cs series from these MD results.

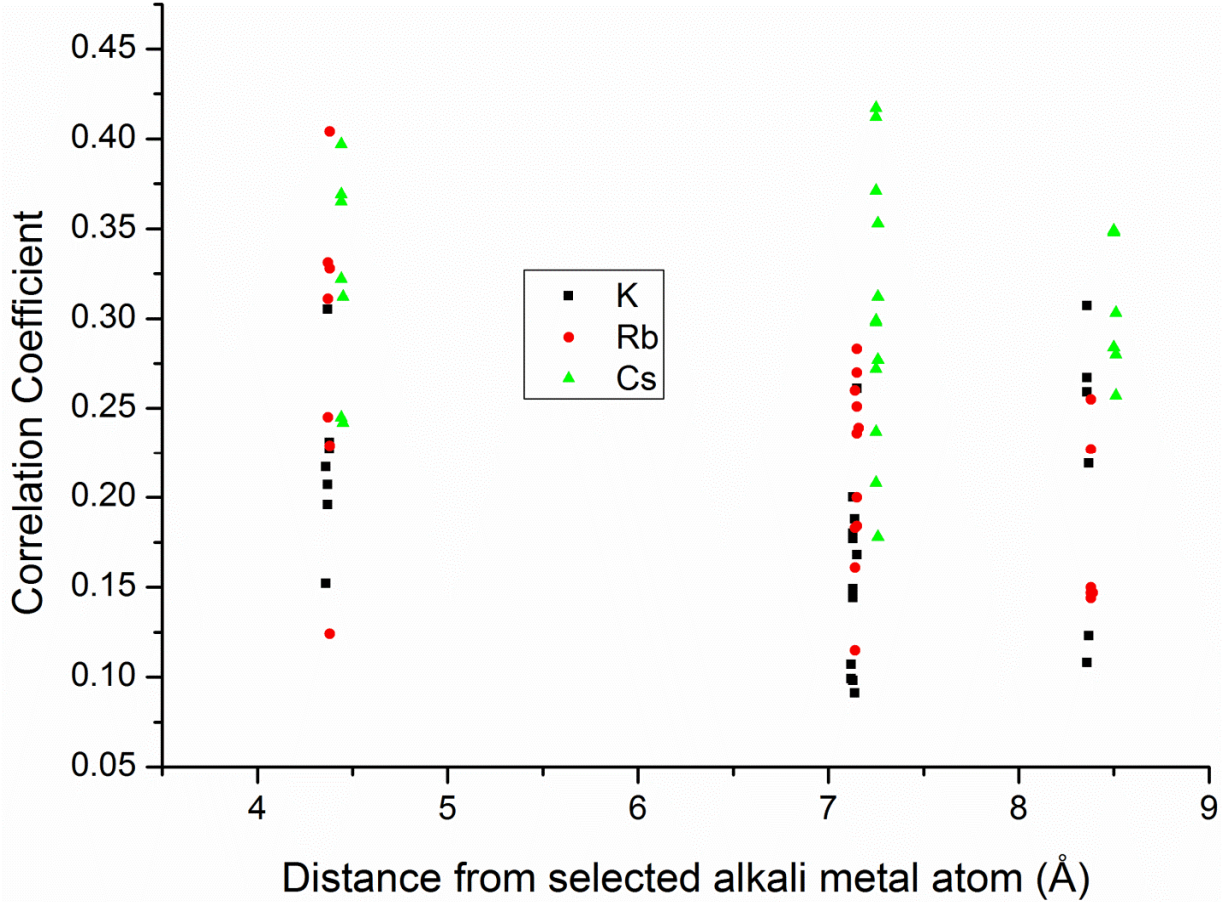

FIG. 6. Trends in the correlation coefficient between the alkali metals as a function of the separation distance. Overall, the correlations increase with atomic number of the alkali metal reflecting the stronger interaction possible at the effectively shorter distances of the larger atoms (see text.). Correlations at distances > 5 Å may be affected by periodic boundary conditions applied in the simulation and are included here only for comparisons across the K-Cs series (see text).

In FIG. 6, only the correlation coefficients on the alkali-metal sublattice are plotted together where it can be seen that the average coefficients increase with atomic number of the alkali metal. We attribute this increase in the correlation coefficients to the size of the metal atom; in these pyrochlores, as the alkali-metal atom gets larger, the ratio of the atom crystal diameter[45] (twice the crystal radii) to the internuclear distance increases (from 0.70 for K to 0.82 for Cs) so that the larger atoms are effectively closer to each other leading to a stronger interaction. For a given alkali metal, the correlation coefficient appears relatively insensitive to the separation distance suggesting the existence of long-range ordering on the alkali-metal sublattice. However, as already noted, conclusions about long-range correlations beyond 5 Å cannot be definitively established from these results because of potential artifacts from the periodic boundary conditions applied to the simulation cell. A larger simulation cell is required in order to conclusively extract the long-range correlations and a study to achieve this is in progress.

Nevertheless, our present results suggest significant correlated motion on the alkali-metal sublattice which may be consistent with Coulombic interactions as previously suggested from the study of an effective ionic Hamiltonian.[9] We note, however, that our results are in contradiction with the conclusion from that study that the K exhibits the strongest correlations. This disagreement could be a result of the different temperatures used in these studies; our results are for 300 K whereas the previous study was at 0 K. Our results suggest that some important features of the alkali-metal dynamics can be understood by restricting the analysis to the alkali-metal sublattice.

### B. The mechanism of rattler coupling in the K, Rb, and Cs series at 300 K

In Sec. III.A we show that the alkali-metal dynamics are more correlated to the dynamics on the alkali-metal sublattice than to those occurring in the cage framework. Significantly, the coupling to the O and Os atoms is quite similar whereas that to the other alkali metals is different. This is evidence that the dynamical coupling between an alkali-metal atom and a cage atom, and that between alkali-metal atoms may be distinct in nature. In this section, we further examine these two different types of coupling in order to clarify the atomistic picture of rattler coupling in these compounds. Since MD offers the option of keeping selected atoms fixed at their initial positions throughout a simulation, we exploit this capability in order to explore the differences between these types of coupling. Performing static lattice simulations permits elimination of the effects of the cage modes from the alkali-metal dynamics. With coupling to the cage modes excluded, it is then possible to isolate and examine the alkali-metal sublattice coupling independently. Figures 7-9 present the alkali-metal MD spectra calculated in nMOLDYN[46] applying the incoherent approximation[47] and in line with our previous work,[11] we calculated the generalized density of states (GDOS) of the vibrational modes by scaling of the density of states (DOS) with the energies. As per this previous work,[11] we noted that the incoherent approximation was inadequate for calculating the simulated INS spectra of the O atoms in these compounds. However, as this approximation works well for the alkali metal atoms, we employ it in this study where the focus is on the alkali-metal dynamics only.

Following the literature on the β-osmates, we discuss rattler dynamics in terms of the $T_{1u}$ and $T_{2g}$ modes which are the irreducible representations (irreps) for a rattler in octahedral ($O_h$) site symmetry. However, we note that the literature is somewhat unclear on the exact symmetry for describing rattler dynamics in these β-pyrochlores. The $O_h$ point group is the crystallographic symmetry for a rattler at the 8b site which is the center of an octahedron formed by the six nearest-neighbor O atoms. For this reason, discussions of rattler dynamics in Raman (which only observes the $T_{2g}$ mode) and neutron spectroscopy have used the $O_h$ irreps.[1, 23, 24, 27, 48] In contrast, most theoretical models have assumed tetrahedral ($T_d$) symmetry for the rattler,[9, 10, 31, 49] where the rattler modes are then discussed in terms of the $A_1$ and $T_2$ irreps of this point group. If alkali-metal atoms occupy the 8*b* sites, then the alkali-metal sublattice forms a diamond lattice giving $T_d$ local site symmetry for the rattler with respect to this sublattice. The main rationale for applying $T_d$ symmetry, which is only the symmetry on the alkali-metal sublattice and not that of the full lattice is that, in these pyrochlores, rattler-rattler interaction is stronger than rattler-cage interaction as already discussed in Sec. III.A. That rattler spectral features calculated from theoretical models assuming $T_d$ symmetry show reasonable agreement with experiment may suggest that $T_d$ is the stronger symmetry for representing the dynamics than the crystallographic $O_h$ symmetry. Although we use the $O_h$ irreps for the rattler modes, we emphasize that further work is needed to clarify this issue. Hasegawa *et al*.[23] and Schoenes *et al*.[48] calculated the phonon modes at the Γ-point but the eigenvectors of the rattler modes were not reported in those contributions. We note that, in analyzing the Q dependence of the $RbOs_2O_6$ response at 300 K, Mutka *et al.*[27] found a strong response at Q=1.1 Å$^{-1}$ corresponding to the [111] zone center which they attributed to a high amplitude longitudinal mode with displacement towards the nearest neighbor A atoms.[27] This corresponded to the high energy band which would be the $T_{2g}$ mode and, according to Hasegawa *et al.*, involves half the A atoms vibrating in opposite directions.[24] Mutka *et al.*[27] also found that the low energy band, which would be the $T_{1u}$ phonon, coincided with the $[\frac{3}{2}\frac{3}{2}\frac{3}{2}]$ zone boundary and, according to Hasegawa *et al.*, is a vibration involving all A atoms moving in the same direction[24]. Our own attempts at calculating the $KOs_2O_6$ phonon modes using VASP were unsuccessful as we found some relatively large negative values (up to ~ -30 meV) for a few of the system frequencies.

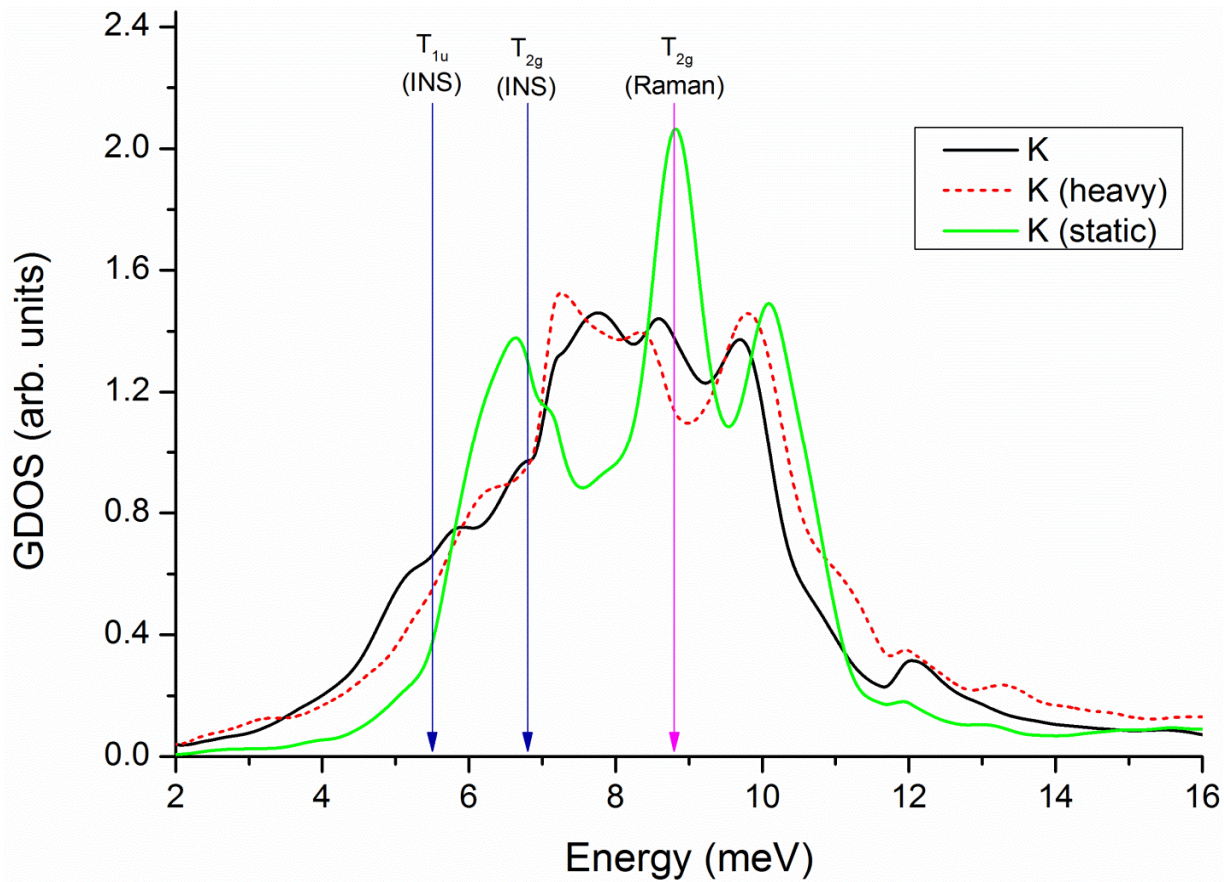

FIG. 7. Spectra of the K element in $KOs_2O_6$ at 300 K calculated from three different MD simulations with the following settings: K – standard simulation, K (heavy) – simulation with the atomic masses of the cage atoms (i.e., O and Os) quadrupled, and K (static) – cage atoms kept fixed at their initial positions. Except for the low-energy shoulder region, the spectrum for the heavy lattice simulation is virtually identical to that of the normal lattice, whereas that of the static lattice shows some distinct features, particularly the sharp peak at low energies. Based on the spectrum obtained for the static lattice, the $T_{1u}$ and $T_{2g}$ modes can be identified more clearly at 6.6 meV and 8.8 meV respectively, the latter coinciding exactly with Raman assignment.[23] The energies of the $T_{1u}$ and $T_{2g}$ modes from INS[1] and Raman scattering[23] are indicated by the arrows.

FIG. 7 shows the calculated K spectra for the standard and static simulations. We have also included a spectrum from a simulation where the masses of the O and Os atoms are increased fourfold, K (heavy). In the harmonic approximation, quadrupling the atomic masses halves the vibrational frequencies which allows assessment of the sensitivity of the coupling between the rattler and cage modes. All the spectra in FIG. 7 exhibit three peaks, and four shoulders (5.2, 5.8, 6.7, and 7.2 meV), one shoulder (6.2 meV), and no shoulder for the standard, heavy, and static spectrum, respectively. The interpretation of these spectra is quite complicated and the key lies in the correct assignment of the low energy peak in the static spectrum at 6.6 meV. The poor agreement between INS and Raman scattering for the $T_{2g}$ mode makes it difficult to decide this peak assignment: Is it a $T_{2g}$ or $T_{1u}$ mode? The INS assignment suggests that it is likely to be the $T_{2g}$ mode leading to the conclusion that the fixed local site-symmetries provided by the static cages around the K atoms result in a broader spectrum with significantly more splitting of the $T_{2g}$ manifold. Further, this interpretation suggests that the $T_{1u}$ mode is not excited in the static cage simulation. However, this interpretation does not seem to be fully consistent with the data. Firstly, the INS $T_{2g}$ assignment coincides with the shoulder at 6.8 meV completely missing the three-peak structure of the standard spectrum. Our calculated spectrum shows much better agreement with the Raman[23] and micro-Raman[48] scattering assignments of the $T_{2g}$ mode which suggests that the peak at 6.6 meV should correspond to the $T_{1u}$ mode. Additionally, features in the low-energy shoulder region suggest that as the cage framework gets less mobile, the system transitions from multiple shoulders to a single shoulder (K (heavy)) which eventually becomes a well-defined peak in the fixed cage case. Secondly, the Rb results (see below) show a clear $T_{1u}$ mode in the static spectrum where the standard spectrum exhibits a shoulder. These considerations weigh in favor of the $T_{1u}$ assignment of the peak at 6.6 meV and we adopt this assignment in the rest of this discussion. If this assignment of the $T_{1u}$ mode is correct, it follows that this mode is significantly more sensitive to the cage dynamics compared to the $T_{2g}$ mode. Figure 7 shows, strikingly, that compared to the static lattice, the intensity of the $T_{1u}$ mode almost vanishes when the cage dynamics are turned on in the normal lattice simulation. It has been reported that the $T_{1u}$ mode softens more strongly with cooling compared to the $T_{2g}$ mode[27, 50, 51] and it is possible that the sensitivity of the $T_{1u}$ mode to the cage dynamics we find here may partly explain that behavior. In our work on $KAl_{0.33}W_{1.67}O_6$, the Al-doped W analogue of $KOs_2O_6$, we found significant hardening of the K phonons with increasing temperature due to a tightening of the highly anharmonic potential around the K atoms in the direction of maximum displacement, driven by entropic effects of the cage.[52] Turning to the $T_{2g}$ mode, we notice that the main spectral features of this mode are somewhat preserved on fixing the cage framework. This suggests that the $T_{2g}$ mode should primarily mediate the K-K coupling on the alkali-metal sublattice.

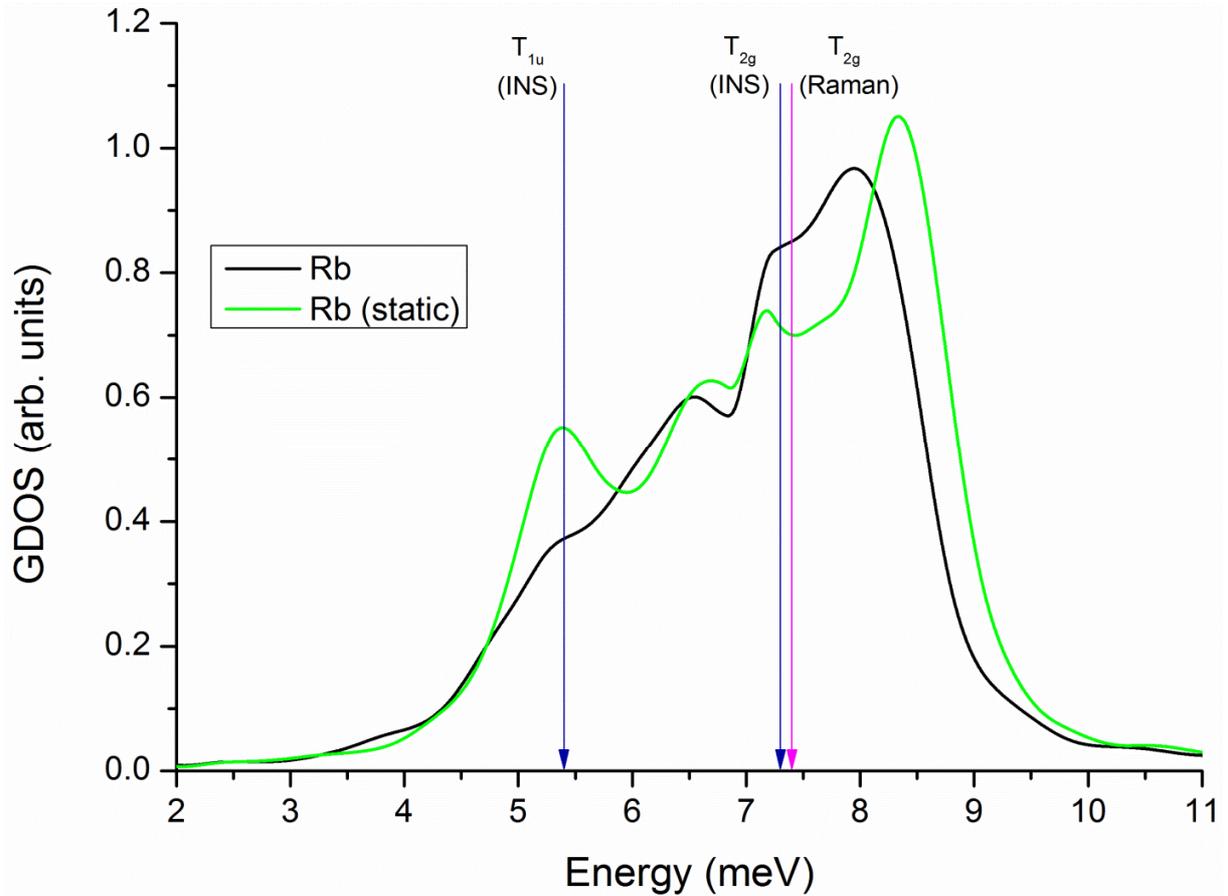

FIG. 8. The vibrational spectra of Rb in RbOs$_2$O$_6$ calculated from 300 K MD simulations. The Rb (static) spectrum is obtained for an MD simulation where all the O and Os are kept in their initial positions throughout the simulation. Indicated by the arrows are the energies of the T$_{1u}$ and T$_{2g}$ modes from INS[27] (as cited in Hiroi et al.[1]), and T$_{2g}$ mode from Raman scattering.[23] The main difference between the spectra is the emergence of a more pronounced T$_{1u}$ peak in the 'static' spectrum which is only a shoulder in the standard spectrum.

Figure 8 shows that the main effect of fixing the cage framework in RbOs$_2$O$_6$ is the emergence of a distinct T$_{1u}$ peak where a shoulder existed while the T$_{2g}$ peak exhibits no major change. That the T$_{1u}$ mode which is distinct in the static lattice spectrum reduces to a shoulder in the standard simulation implies that this mode plays a significant role in the coupling between the Rb rattler and the cage modes. On the other hand, the insensitivity of the T$_{2g}$ mode to the fixing of the lattice suggests that this mode may play no crucial role in the coupling between the Rb rattler and the lattice. Instead, as in the K case, the results suggest that the Rb T$_{2g}$ mode is responsible for the rattler-rattler coupling occurring on the Rb sublattice. This is a significant clarification of the mechanism of rattler coupling in RbOs$_2$O$_6$. The results for Cs, FIG. 9, present the opposite picture; here the distinct T$_{1u}$ peak exhibited by the standard spectrum is absent from the static spectrum, implying that for Cs, this mode is excited by coupling to the cage modes. This result is somewhat surprising; although it still implies that the T$_{1u}$ mode is important for coupling to the cage modes as in both K and Rb, it makes the Cs $T_{1u}$ mode distinct from the K and Rb cases in its mode of excitation. Also evident from FIG. 9 is that the Cs $T_{2g}$ mode becomes a single sharp peak when the cage modes are eliminated by fixing the cage framework implying that the splitting in this peak in the standard spectrum may be a consequence of some coupling to the cage modes. This is consistent with FIG. 4-6 which show that Cs exhibits the strongest correlations to the cage dynamics in the K-Cs series. However, the change in the $T_{2g}$ is not as pronounced as that of the $T_{1u}$ mode indicating that this mode couples the Cs rattler to other Cs atoms more than it couples it to the lattice.

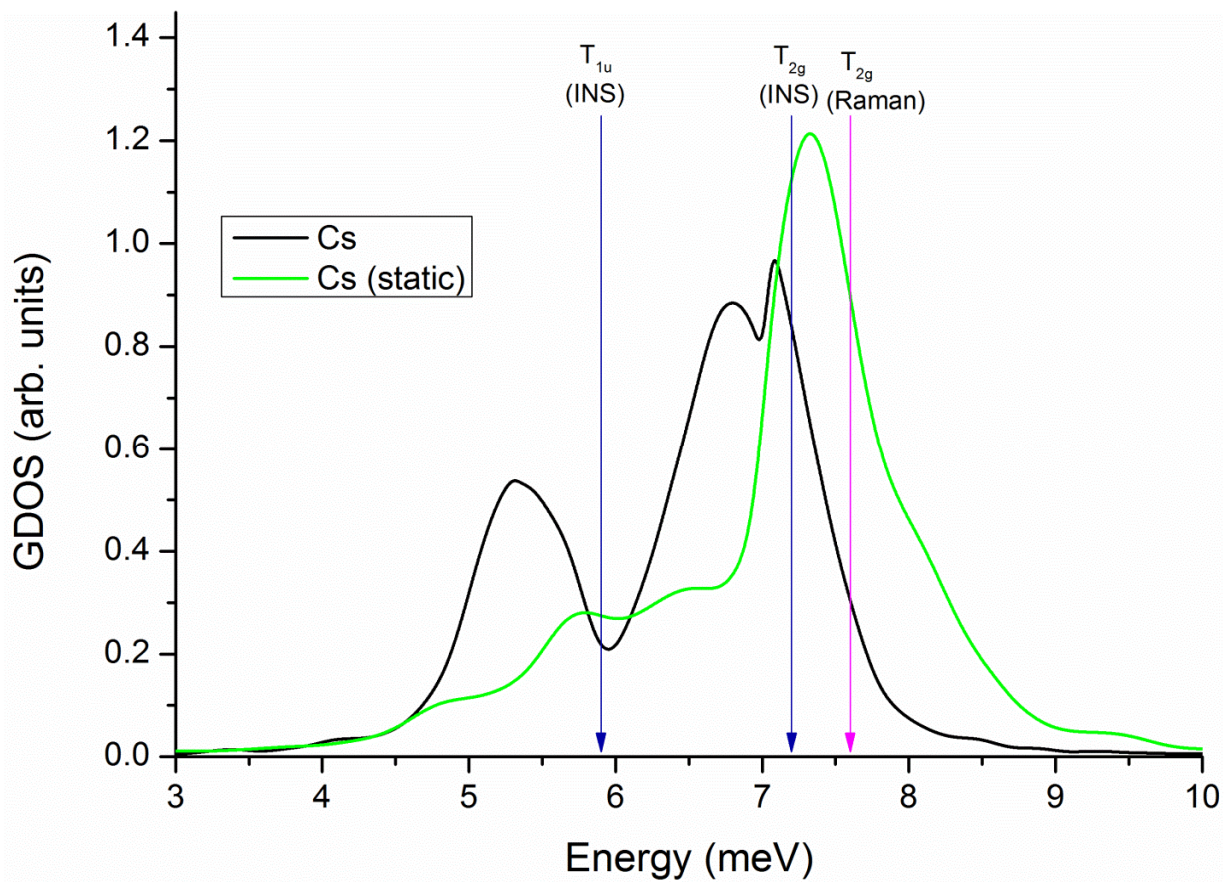

**FIG. 9. MD spectra of the Cs modes at 300 K. The Cs (static) spectrum is calculated from a simulation in which the O and Os atoms are kept fixed at their initial positions throughout the simulation. The key difference between these spectra is the missing $T_{1u}$ mode from the static lattice spectrum indicating that this mode is excited by coupling to the cage modes in $CsOs_2O_6$.**

Although the exact details are not simple and are different for the different pyrochlores, the overall picture that emerges from FIG. 7-9 is that the $T_{1u}$ mode couples the rattler more strongly to the cage modes than it couples it to other rattlers. The $T_{2g}$ mode is the exact opposite, coupling the rattlers to each other more strongly than to the cage modes. This result is a crucial clarification to our understanding of the rattler dynamics in these beta pyrochlores and it means that the general statement that the alkali-metal modes do not couple to the lattice modes may not be completely accurate. Perhaps a more precise statement is that the $T_{1u}$ mode significantly couples to the lattice modes while the $T_{2g}$ coupling is weaker; but because the latter represents the dominant dynamics, it determines the main features observed for the alkali-metal dynamics in Raman scattering, INS, and MD work. From these results, it is likely that the dynamical characteristics of an alkali-metal atom in the β-osmates are mainly determined by the strength of these two quite distinct types of coupling; the $T_{1u}$ mode to the cage modes and the $T_{2g}$ mode to other alkali-metal atoms. This means that it should then be possible to explain the differences in alkali-metal dynamics observed along the K-Cs series on the basis of these two types of coupling although the exact details remain quite complicated. Further work needs to be done to clarify the $T_{1u}$ and $T_{2g}$ mode assignments in K as well questions regarding the local symmetries at the alkali-metal sites which cannot be established definitively from these relatively short simulations.

### C. Spectral broadening in the K-Cs series

It is evident from a comparison of the spectra from both the standard and static simulations in FIG. 7-9 that there is considerable broadening of these spectra from Cs to K. In a recent contribution[11], we discussed this phenomenon and showed that this could be partly understood from the flattening of the local potential around these atoms from Cs to K. We extend this explanation to include the role of the correlated motion on the alkali-metal sublattice. By examining the dynamics of the immediate environment of the alkali metal based on its sublattice, we expect to be able to find an explanation for the trend in the broadening of the spectra in FIG. 7-9. Since the diamond alkali-metal sublattice presents each A ion with a $T_d$ local symmetry, in principle, one could perform a group theoretic treatment of the MD trajectories to decompose them into the irreps of this point group. Such an analysis will be insightful as it enables the precise identification of the different normal modes which could clarify the issue of the $T_{1u}$ and $T_{2g}$ irreps discussed earlier. However, our attempt at this was unsuccessful and the reason appears to be that a 36 ps simulation may not be long enough to give the correct time-average local symmetry at each site. Instead, in what follows we consider the dynamics of the volume of the general tetrahedron whose vertices are the coordinates of the nearest neighbors of a selected alkali-metal site. For each pyrochlore, we calculated the volume at each simulation step around the selected site and performed a (forward) fast Fourier-transform (FFT) of this function to obtain volume fluctuation spectra. The statistics of the volume fluctuations are summarized in Table I while the corresponding spectra are plotted in FIG. 10. The K statistics in Table I stand out with the volume fluctuating over the widest range giving a standard deviation 35 % larger than Rb and Cs.

Table I. Summary statistics of the volume of the general tetrahedron (Å$^3$) formed by the nearest neighbors of a site on the alkali-metal sublattice. The results show that the K system has the largest variation in the volume reflecting the greater range of motion for the smaller K atoms.

| Pyrochlore | Mean | Minimum | Maximum | Standard Deviation |
|---|---|---|---|---|
| K | 42.67 | 35.47 | 50.31 | 2.60 |
| Rb | 43.00 | 37.76 | 48.61 | 1.93 |
| Cs | 45.02 | 40.51 | 50.51 | 1.93 |

The gross features of the spectra in FIG. 10 correspond with those in FIG. 7-9 quite well, clearly exhibiting the general increase in broadening from Cs to K. Since similar broadening is obtained for static lattice calculations, it suggests that the dynamics responsible for the spectral broadening primarily occur on the alkali-metal sublattice.

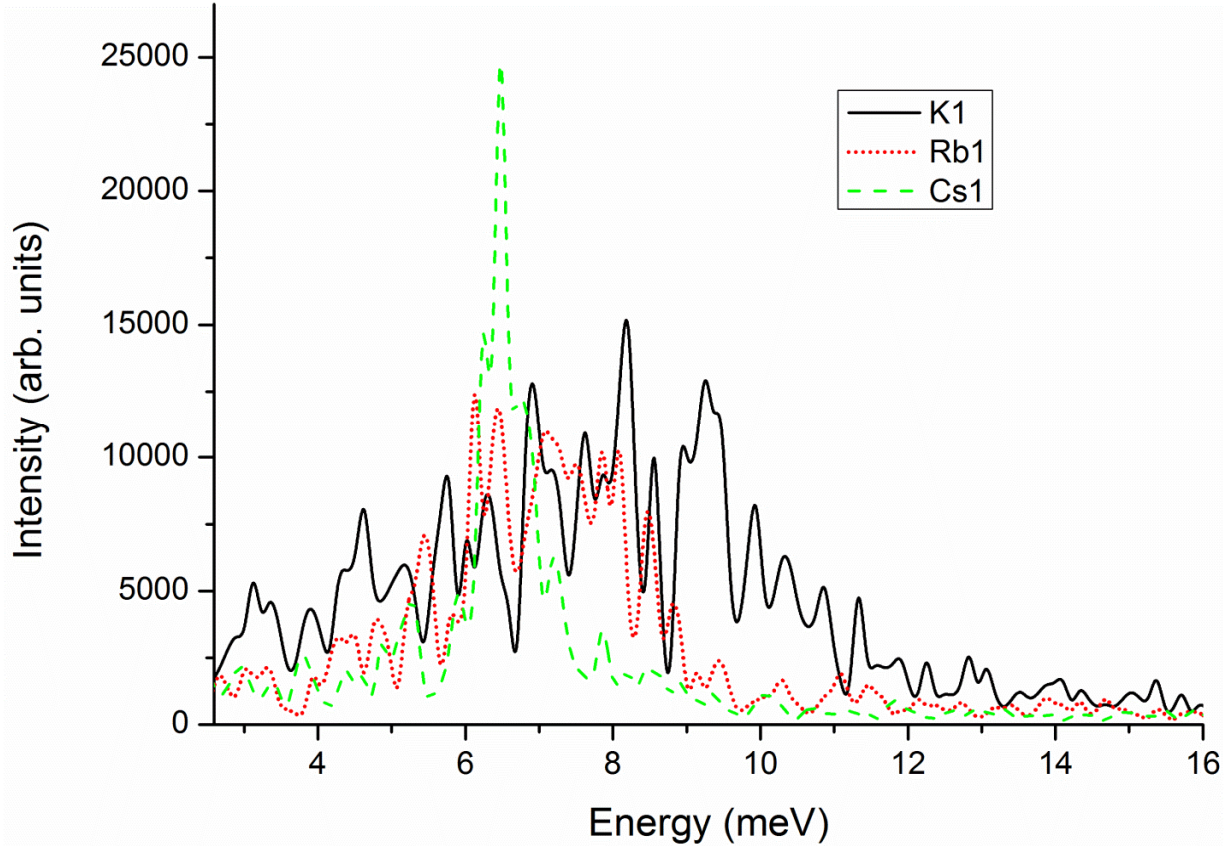

FIG. 10. Frequency spectra for the volume fluctuations of the tetrahedra whose vertices are the nearest neighbors of a selected alkali-metal site (K1, Rb1, or Cs1) on the alkali-metal sublattice calculated from the MD simulation. The range of vibrational frequencies increases with decreasing size of the alkali metal indicating that the smaller the atom, the greater is the range of motions it can undergo.

To get a geometric picture of the types of distortions occurring in the tetrahedra leading to the dynamics of FIG. 10, we calculated six-dimensional correlation matrices between the edges of the tetrahedra (whose vertices are alkali-metal sites) and the results are summarized in Table II. The schematics in FIG. 11 are based on Table II and the arrows on the edges represent the correlation coefficients. We represent a positive correlation by two outward pointing arrows and the converse is true for a negative correlation. Correlations of magnitude ≤ 0.1 are designated with a star. The lengths of these arrows are roughly proportional to the magnitudes of the correlation coefficients they represent. Two main observations can be made from the results both in Table II and FIG. 11. Firstly, overall, the correlation coefficients for the K are comparable to those for Rb, although somewhat lower, whereas those for Cs are generally larger. The larger correlation coefficients for Cs suggest a more rigid lattice and, hence, more restricted ranges of motion for these atoms compared to K and Rb. Secondly, of the fifteen distinct correlation coefficients calculated for each pyrochlore, three pairs of edges are anti-correlated for both K and Rb whereas there are only two for Cs. Using the notation in FIG. 11, these edges are 5-6/7-8, and 5-7/6-8 for K, Rb, and Cs as well as 5-8/6-7 for K and Rb only. Note that the designation of correlation coefficients of magnitude ≤ 0.1 by a star in Figure 17 obscures this for the 5-8/6-7 case of K. Anticorrelation (negative coefficient) adds flexibility to the dynamics of the system as (positive) correlation alone only produces the 'breathing mode' of the tetrahedron. Consequently, the lower anticorrelation in the Cs dynamics implies reduced freedom of motion in this system. Further, although both K and Rb have the same number of anti-correlated dynamics, the correlation coefficients of K for these dynamics are much lower suggesting even greater flexibility of motion for this system. We therefore conclude that the broadening of the alkali-metal spectra from Cs to K in FIG. 7-9 arises from increasing flexibility of motion with decreasing atomic size. The features in the K spectrum make it plausible to suggest that the K sites couple to each other through a larger variety of vibrational motions thanks to the smaller size of the K atom which permits greater flexibility of motion, the latter leading to a lower average site symmetry. Thus it appears, from this analysis, that the interaction that matters most is the one on the alkali-metal sublattice and, perhaps counterintuitively, the stronger this interaction is, the sharper the spectrum.

Table II. Six-dimensional correlation matrix for the edges of the tetrahedron whose vertices are the nearest-neighbors of site A1 (A = K, Rb, Cs) in the MD simulation cell. The edges are labeled as Ai-Aj (A = K, Rb, Cs; i,j = 5,6,7,8) where Ai-Aj is the distance between sites Ai and Aj. Each row (or column) represents a distinct vibrational mode for the tetrahedron of nearest-neighbors.

| $KOs_2O_6$ | | | | | | |
|---|---|---|---|---|---|---|
| | K5-K6 | K5-K7 | K5-K8 | K6-K7 | K6-K8 | K7-K8 |
| K5-K6 | 1.00 | 0.19 | 0.13 | 0.20 | 0.16 | -0.13 |
| K5-K7 | 0.19 | 1.00 | 0.23 | 0.17 | -0.12 | 0.09 |
| K5-K8 | 0.13 | 0.23 | 1.00 | -0.08 | 0.22 | 0.27 |
| K6-K7 | 0.20 | 0.17 | -0.08 | 1.00 | 0.11 | 0.24 |
| K6-K8 | 0.16 | -0.12 | 0.22 | 0.11 | 1.00 | 0.37 |
| K7-K8 | -0.13 | 0.09 | 0.27 | 0.24 | 0.37 | 1.00 |

| $RbOs_2O_6$ | | | | | | |
|---|---|---|---|---|---|---|
| | Rb5-Rb6 | Rb5-Rb7 | Rb5-Rb8 | Rb6-Rb7 | Rb6-Rb8 | Rb7-Rb8 |
| Rb5-Rb6 | 1.00 | 0.01 | 0.31 | 0.13 | 0.21 | -0.28 |
| Rb5-Rb7 | 0.01 | 1.00 | 0.19 | 0.12 | -0.33 | 0.17 |
| Rb5-Rb8 | 0.31 | 0.19 | 1.00 | -0.21 | 0.27 | 0.13 |
| Rb6-Rb7 | 0.13 | 0.12 | -0.21 | 1.00 | 0.20 | 0.29 |
| Rb6-Rb8 | 0.21 | -0.33 | 0.27 | 0.20 | 1.00 | 0.16 |
| Rb7-Rb8 | -0.28 | 0.17 | 0.13 | 0.29 | 0.16 | 1.0000 |

|         | CsOs$_2$O$_6$ | | | | | |
|---------|---------|---------|---------|---------|---------|---------|
|         | Cs5-Cs6 | Cs5-Cs7 | Cs5-Cs8 | Cs6-Cs7 | Cs6-Cs8 | Cs7-Cs8 |
| Cs5-Cs6 | 1.00    | 0.07    | 0.23    | 0.28    | 0.37    | -0.16   |
| Cs5-Cs7 | 0.07    | 1.00    | 0.30    | 0.29    | -0.07   | 0.44    |
| Cs5-Cs8 | 0.23    | 0.30    | 1.00    | 0.10    | 0.22    | 0.45    |
| Cs6-Cs7 | 0.28    | 0.29    | 0.10    | 1.00    | 0.18    | 0.22    |
| Cs6-Cs8 | 0.37    | -0.07   | 0.22    | 0.18    | 1.00    | 0.13    |
| Cs7-Cs8 | -0.16   | 0.44    | 0.45    | 0.22    | 0.13    | 1.00    |

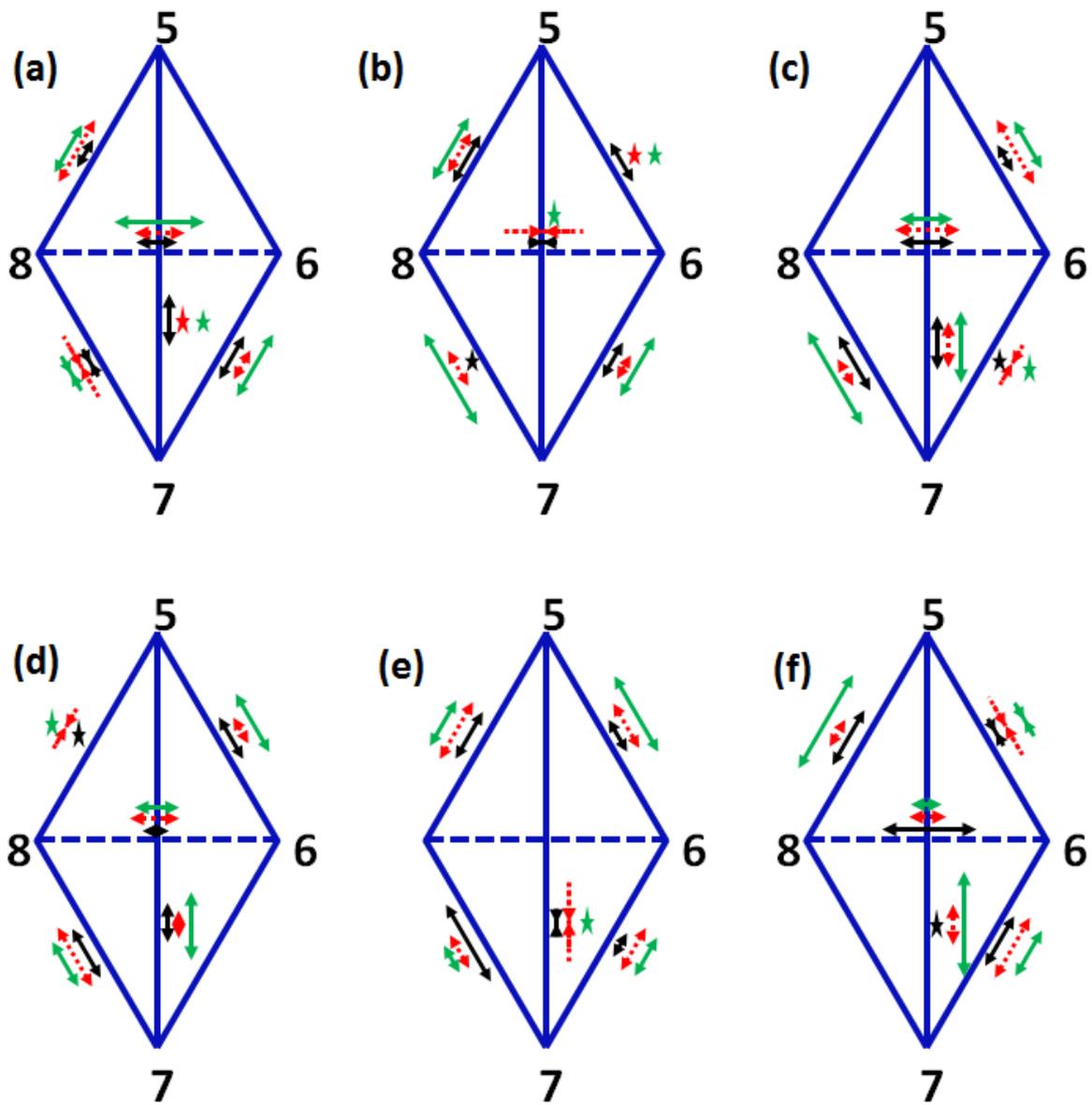

FIG. 11. Schematic of the tetrahedral dynamics showing the complex distortions of the tetrahedra enclosing K1, Rb1, and Cs1 in the K, Rb, and Cs pyrochlores, respectively. The figures a, b,… correspond to rows (or columns) 1, 2,… in Table II, respectively. The vertices of the tetrahedra correspond to the alkali-metal sites in the MD simulation cell with the labels matching the site labels. Correlation coefficients are plotted on the each edge of the tetrahedra as arrows where double outward (inward) arrows represent positive (negative) correlation. For each edge, the arrows are for K (black), Rb (red), and Cs (green) moving away from the edge of the tetrahedron. In each case, the unmarked edge is the edge being stretched, i.e., correlation coefficient of unity. All but the 5-7 edge are the same length which is defined as a correlation of unity for calculating the lengths of the correlation arrows. Because of the difficulty of representing small lengths, correlation coefficients of magnitude ≤ 0.1 are represented by stars which, unfortunately, obscure some important features for the K system – see text. The broad feature of increasing correlation coefficients with atomic size suggests a corresponding increase in the 'rigidity' of the sublattice which limits flexibility of motion.

Our results have important implications: Firstly, the strong coupling we find on the alkali-metal sublattice indicates that important aspects of the rattling dynamics in these materials can be understood from a restricted analysis considering only the alkali-metal sublattice. This could prove very advantageous for studies involving MD simulations and effective model Hamiltonians, as in both cases, the system size could be reduced considerably. For instance, in the case of MD simulations, it may be possible to run simulations on the alkali-metal sublattice only. As the resulting unit cell is only 1/9 of the normal cell (based on total number of atoms), it becomes feasible to implement larger supercells which would permit the extraction of long-range properties from these simulations. We have successfully optimized the geometry of a 2x2x2 supercell for Rb and an MD simulation will be run to confirm this prediction. If this proves feasible, then this simpler system offers the possibility of providing insight into the role of the conduction electrons from the alkali metals in superconductivity. Secondly, the existence of strong coupling on the alkali-metal sublattice means that significant tuning of these dynamics could be achieved by targeted modifications of this sublattice. This will be particularly true for the Cs pyrochlore. Thirdly, the synthesis of the $LiOs_2O_6$ and $NaOs_2O_6$ remains a challenge. As the A-A coupling weakens with decreasing atomic mass of the alkali metal, it suggests that this sublattice coupling may account for a significant part of the stabilization energy of the crystals of these compounds. If this is the case, it becomes plausible to expect that both $LiOs_2O_6$ and $NaOs_2O_6$ could form under high (chemical or mechanical) pressure when the ratio of the Li-Li (or Na-Na) internuclear distance to the Li (or Na) crystal diameter is at least equal to that of $KOs_2O_6$. Fourthly, although this study is at a much higher temperature (300 K) than the reported $T_c$ s of these materials, it provides clues as to why K might have the highest $T_c$ of these pyrochlores. The flexibility in the types of motion exhibited by the K (broad range of vibrational frequencies, FIG. 10) suggests there is a greater chance for some of these motions to be selected to couple to the electronic degrees of freedom of the system leading to conventional superconductivity. Lastly, the identification of the somewhat distinct roles of the $T_{1u}$ and $T_{2g}$ modes in rattler coupling is a major step towards the development of a more complete mechanism of rattler coupling in the β-pyrochlores.

## IV. CONCLUSION

We have shown from *ab initio* MD that alkali-metal atoms (rattlers) in the β-pyrochlore osmates couple to each other more strongly than they do to the cage atoms. The MD evidence indicates that this rattler-rattler coupling is mediated through the $T_{2g}$ mode whereas the weaker coupling to the cage occurs through the $T_{1u}$ mode. The $T_{2g}$ mode controls the main features of rattler dynamics, showing that the trend in spectral broadening from Cs to K is a consequence of the K vibrating at a range of closely-spaced frequencies due to the ability of this relatively small rattler to undergo a broader range of motions. The identification of the somewhat distinct roles of the $T_{1u}$ and $T_{2g}$ modes in rattler coupling provides an important stepping stone towards developing a more complete microscopic mechanism of rattler coupling, not only in these osmates, but also for a broader class of similar materials including clathrates and skutterudites. Calculation of the phonon eigenvectors for the K atoms in $KOs_2O_6$, even if limited only to the Γ-point, will clarify the irreps of the K modes and facilitate their assignment.

# ACKNOWLEDGEMENTS

This work was partly supported by the Multi-modal Australian ScienceS Imaging and Visualisation Environment (MASSIVE) (www.massive.org.au).